\documentclass[preprint,12pt]{elsarticle}

\usepackage{amssymb}

\journal{Annals of Physics}

\begin{document}

\begin{frontmatter}



\title{Novel symmetries in ${\cal N} = 2$ supersymmetric quantum mechanical models}


\author{R. P. Malik\corref{cor1}}
\ead{malik@bhu.ac.in}
\address{Physics Department, BHU-Varanasi-221 005, India\\
and\\
DST-CIMS, Faculty of Science, BHU-Varanasi-221 005, India}

\author{Avinash Khare}
\ead{khare@iiserpune.ac.in}
\address{Indian Institute of Science for Education and Research, Pune-411 021, India}
\cortext[cor1]{Corresponding author}

\begin{abstract}
We demonstrate the existence of a novel set of discrete symmetries in the context 
of ${\cal N} = 2$ supersymmetric (SUSY) quantum mechanical model with a potential 
function $f(x)$ that is a generalization of the  potential of the 1D SUSY harmonic 
oscillator. We perform the same exercise for the motion of a charged particle in 
the $X-Y$ plane under the influence of a magnetic field in the $Z$-direction. We 
derive the underlying algebra of the {\it existing} continuous symmetry transformations 
(and corresponding conserved charges) and establish its relevance to the  algebraic 
structures of the de Rham cohomological operators of differential geometry. We show 
that the discrete symmetry transformations of our present general theories correspond 
to the Hodge duality operation. Ultimately, we conjecture that any arbitrary 
${\cal N} = 2$ SUSY quantum mechanical system can be shown to be a tractable model 
for the Hodge theory.
\end{abstract}


\end{frontmatter}


\section{Introduction}
\label{}

The supersymmetric (SUSY) quantum mechanical models represent mathematically one of the most
beautiful and elegant examples in the realm of theoretical physics which have found 
applications in diverse domains of physical phenomena (see, e.g. [1,2]). 
At the level of quantum mechanics, supersymmetry
connects two Hamiltonians and corresponding states and, at the classical level, this
symmetry transforms the commuting dynamical variables into the anticommuting ones and vice-versa. 
The central theme of our present investigation is to explore new set of continuous and
discrete symmetries of totally different kinds  of SUSY models and demonstrate that these
symmetries provide physical realizations of abstract properties
associated with the cohomological operators of differential geometry [3-5]. As a result, we conjecture
that ${\cal N} = 2$ SUSY models, obeying $sl(1/1)$ superalgebra, belong to a
special class which can be shown to be physical models for the Hodge theory.

From the physical point of view, the above kind of studies are very important. For instance,
in our earlier series of research works [6-10], we have established that the Abelian 1-form, 2-form and
3-form gauge theories (in 2D, 4D and 6D dimensions of spacetime) are models for the Hodge theory
within the framework of Becchi--Rouet-Stora-Tyutin (BRST) formalism. Furthermore, the (non-)Abelian
1-form gauge theory in 2D [6], interacting 2D Abelian 1-form theory with Dirac fields [7], 4D
free Abelian 2-form gauge theory [8], 6D free Abelian 3-form gauge theory, etc., are endowed with continuous
and discrete set of symmetry transformations that provide physical realizations for the de Rham
cohomological operators, Hodge duality operation, degree of a form, etc., within the purview of
BRST formalism. The culmination of all the above studies is the proof that the 2D
(non-)Abelian gauge theories (without any interaction with matter fields) belong to a new 
class of topological field theories [11] and 4D free Abelian 2-form as well as 6D free Abelian 3-form
gauge theories turn out to be
the models for quasi-topological field theory [12,10].

All the above cited theoretical models, however, belong to a special class of field theories 
(i.e. gauge theories) that are endowed with first-class constraints in the terminology of Dirac's prescription
for the classification scheme [13]. In a very recent paper [14], we have taken a one (0 + 1)-dimensional (1D)
${\cal N} = 2$ SUSY model of harmonic oscillator and shown that it provides a physical model for the Hodge theory.
In  our present endeavor, we consider two different kinds of physically interesting
${\cal N} = 2$ SUSY models 
and demonstrate that they respect all the pertinent symmetries that provide a basis for these physical systems
to be the models for Hodge theory because all the cohomological operators (and connected Hodge duality
operation) find their physical realizations in terms of continuous and discrete symmetry transformations
of the theory. 
Furthermore, the conserved charges, obeying the $sl(1/1)$ superalgebra, are found to be the analogues of 
de Rham cohomological operators of differential geometry from various points of view. As a consequence, 
we conjecture that the specific class of models, corresponding to
${\cal N} = 2 $ SUSY theories, represent physical models for the Hodge theory.

In our present endeavor, we discuss explicitly the continuous fermionic symmetry
transformations (corresponding to ${\cal N} = 2$ supersymmetry) and derive the corresponding
supercharges by exploiting the Noether's theorem. We also derive the conserved charge corresponding to
a bosonic symmetry that is an anticommutator of the above two SUSY transformations. As expected, we observe that
this bosonic symmetry transformation turns out to be equivalent to a time translation. This observation is
sacrosanct for any well-defined SUSY theory where it is a crucial requirement that two successive SUSY
transformations must produce the spacetime translations in a given dimension of spacetime. In our present
couple of systems (corresponding to  ${\cal N} = 2$ SUSY models), 
the generator of the bosonic symmetry transformations turns out
to be connected with the Hamiltonian of the theory because, as is well-known, the latter is the generator
of the time translation.

Our present investigation is essential on the following counts. First, we have shown, in our
earlier work [14], that the 1D SUSY harmonic oscillator provides a prototype example of a Hodge theory. 
Thus, it is very tempting to study other ${\cal N} = 2$ SUSY models and check whether they also respect
similar kinds of symmetries as does the SUSY oscillator. Second, the motion
of an electrically charged particle under influence of an EM field is a physically very important
topic. Thus, its ${\cal N} = 2$ SUSY quantum mechanical version is interesting in its own right.
To say something {\it new} about this model is always challenging. We show, in our
present endeavor, that this system, too, is a tractable model for the Hodge theory. Finally,
we go a step further and conjecture  that any arbitrary ${\cal N} = 2$ SUSY quantum mechanical model
would be endowed with symmetries that would turn out to be the realizations of cohomological operators.
As a consequence, these SUSY systems represent a special class of models that provide a realization
of Hodge theory.

Besides the above motivations, our present study of simple SUSY quantum mechanical systems would
provide insights into the understanding of ${\cal N} = 2$ SUSY gauge theories of phenomenological 
importance where the cohomological structure might appear. As a consequence, one would be able to
apply the celebrated Hodge decomposition theorem in defining the {\it physical state} of the theory (which
would be chosen to be the harmonic state). The latter would be, naturally, annihilated by the operator
form of $Q, \bar Q$ and $W$ [cf.(16)]. This would put constraints on the theory which will be useful
in the counting of degrees of freedom of the theory. This information would enable
us to study the topological nature of the SUSY gauge theory. In fact, the fermionic charges $Q$ and
$\bar Q$ would play important roles in expressing the Lagrangian density as well as energy-momentum tensor
of such theories. As a consequence, one would be able to state that the energy excitation of the physical
state  would be zero if the physical state is chosen to be the harmonic state in the Hodge decomposition
theorem. Such kind of study has been performed in the context of {\it usual} (non-)Abelian 2D gauge
theories [11].

The material of our present investigation is organized as follows. To set up the notations
and conventions, we start off with a brief synopsis of ${\cal N} = 2$ supersymmetric harmonic
oscillator and discuss its various continuous as well as discrete symmetry transformations in Sec. 2.
Our Sec. 3 is devoted to the discussion of continuous symmetries and the derivation
of corresponding Noether conserved charges for two different ${\cal N} = 2$ supersymmetrical models. 
Our Sec. 4 deals with the discrete symmetries of the above two supersymmetric quantum mechanical systems.
We deduce the algebraic structures of the symmetry operators 
(and corresponding conserved charges) and establish their connection with the algebra of 
cohomological operators in Sec. 5. Finally,
we make some concluding remarks in Sec. 6.

In our Appendix A, we discuss simpler ways of deriving the $sl(1/1)$ closed superalgebra amongst the
conserved charges of ${\cal N} = 2$ SUSY quantum mechanical models that are topics of discussion
in our present endeavor.

{\it Conventions and Notations:} Through out the whole body of our text, the fermionic 
($ s_1^2 = 0, s_2^2 = 0$) symmetries
[that are the analogue of the nilpotent (co-)exterior derivatives] have been denoted by $s_1$ and $s_2$
and their anticommutator (which is an analogue of the Laplacian operator) is represented by
$s_\omega = \{ s_1, s_2 \}$ for all the models of ${\cal N} = 2$ SUSY quantum mechanics. The
corresponding conserved charges have been expressed by $Q, \bar Q, W$. This has been done purposely,
so that, some common features of the above SUSY models could be expressed in a 
concise fashion (see, e.g. Sec. 5 below).

\section{Preliminaries: SUSY oscillator}
We begin with the Lagrangian for a one (0 + 1)-dimensional (1D) supersymmetric harmonic oscillator
which is described by  the ordinary bosonic position variable $x$ and a pair of Grassmannian
variables $(\psi, \bar\psi)$ (with $\psi^2 = \bar\psi^2 = 0, \psi \bar\psi + \bar\psi \psi = 0$)
at the classical level. For the sake of simplicity, we take the mass $m$ of  the oscillator to be
one (i.e. $m = 1$) in the following Lagrangian (with natural oscillator frequency $\omega$) 
(see, e.g. [14] for details)
\begin{eqnarray}
L_0 = \frac{\dot x^2 (t)}{2} - \frac{1}{2} \;\omega^2 \;x^2 (t) + i \;\bar\psi (t) \;\dot \psi (t) 
- \omega \; \bar \psi (t) \;\psi (t),
\end{eqnarray}
where $\dot x = dx/dt$ and $\dot \psi = d\psi / dt$ are the generalized ``velocities'' in terms of the
variation of the instantaneous bosonic and fermionic 
variables $x$ and $\psi$ with respect to the evolution parameter $t$.

The above starting Lagrangian is endowed with the following on-shell nilpotent ($s_1^2 = s_2^2 = 0$) 
infinitesimal symmetry transformations [14]
\begin{eqnarray}
&& s_1 x = \frac{-i \;\psi}{\surd (2 \;\omega)}, \qquad s_1 \psi = 0, \qquad  s_1 \bar \psi =
\frac{1}{\surd (2 \;\omega)} \;(\dot x + i \;\omega \;x ), \nonumber\\
&& s_2 x = \frac{i \;\bar \psi}{\surd (2 \;\omega)}, \qquad s_2 \bar \psi = 0, \qquad  s_2 \psi =
\frac{1}{\surd (2 \;\omega)} \;( - \dot x + i \;\omega \;x ), 
\end{eqnarray}
because the Lagrangian transforms to a total time derivative under $s_1$ and $s_2$. As a consequence,
the action integral ($ S = \int dt \;L_0$) remains invariant under the above continuous  and 
infinitesimal SUSY transformations.

There is yet another continuous symmetry in the theory that is obtained by
taking the anticommutator of the above SUSY transformations $s_1$ and $s_2$ modulo a factor of $i$.
The infinitesimal version of  this bosonic symmetry $s_\omega = \{ s_1, s_2 \}$, for the relevant 
dynamical variables of the theory, are 
\begin{eqnarray}
s_\omega x = \frac{1}{\omega} \;\dot x, \qquad
s_\omega \psi = \frac{1}{2 \;\omega}\; \bigl (\dot \psi - i \;\omega \;\psi \bigr ), \qquad
s_\omega \bar \psi = \frac{1}{2 \;\omega} \;\bigl (\dot {\bar \psi} + i\; \omega \;\bar \psi \bigr ).
\end{eqnarray}
It can be checked that the Lagrangian in (1) transforms to a total derivative under the above
infinitesimal transformations, too, thereby rendering the action integral invariant [14]. 
Thus, ultimately, we have three
continuous symmetries in the theory, out of which, two are fermionic and one is bosonic.

Now we dwell a bit on the existence of a discrete set of symmetries in the theory. These transformations
are responsible for the beautiful connection between the two SUSY continuous symmetries $s_1$ and $s_2$ that
have been discussed above. These explicit and useful discrete transformations are 
\begin{eqnarray}
x \to  - \;x, \quad t \to + t, \quad \omega \to - \;\omega, \quad \psi \to \pm \;i \;\bar \psi, \quad
\bar\psi \to \mp \;i \;\psi,
\end{eqnarray}
under which the Lagrangian (1) transforms to itself (i.e. $ L_0 \to L_0$). Thus, finally, we conclude that there 
are, in totality, five symmetries in the theory. Three of them are continuous in nature and two are discrete.

We note that the SUSY symmetry transformation $s_1$ 
corresponds to the exterior derivative $d$ (with $d^2 = 0$) of differential geometry. 
On the other hand, the nilpotent
($s_2^2 = 0$) SUSY symmetry transformation $s_2$ stands for the co-exterior derivative $\delta$ 
(with $\delta^2 = 0$). This is due to the fact that we have the following 
operator relationships (see, e.g. [14])
\begin{eqnarray}
s_2 \;\Phi = \pm\; *\; s_1 \; *\; \Phi, \quad s_1^2\; \Phi = 0, \quad s_2^2 \;\Phi = 0, \quad \Phi = x, \psi, \bar\psi,
\end{eqnarray}
which mimic the relationship $\delta = \pm * d *, d^2 = \delta^2 = 0$ 
of differential geometry. It should be noted that the
$*$, in the above equation (5), corresponds to the discrete set of symmetries quoted in (4). Thus, the discrete
symmetry transformations (4) stand for the Hodge duality $*$ operation of differential geometry which
connects the (co-)exterior derivatives by: $\delta = \pm * d *$.

Pertinent to the above discussions, we note that the outcome of two successive discrete transformations on
the generic variable $\Phi (t)$ is 
\begin{eqnarray}
*\; [ \; *\; \Phi\;] = +\; \Phi,  \qquad \qquad \Phi = x, \psi, \bar\psi.
\end{eqnarray}
Following the strictures, laid down by the duality invariant theories [15], there would
be only a positive sign in the relationship (5) due to the positive sign present in the above generic
equation (6). Thus, the correct version of (5), consistent with a correct duality-invariant theory, is [15] 
\begin{eqnarray}
s_2 = +\; *\; s_1 \; *, \;\qquad \;s_1^2 = 0, \;\qquad \;s_2^2 = 0,
\end{eqnarray}
As a consequence, only one of the two transformations, listed in (4), would be physically useful.
This can be succinctly expressed as
\begin{eqnarray}
x \to  - \;x, \quad t \to\; +t, \quad \omega \to - \;\omega, \quad \psi \to +\;i \;\bar \psi, \quad
\bar\psi \to - \;i \;\psi.
\end{eqnarray}
To sum up, we have precisely a single discrete symmetry in the theory as given in the above equation.
Thus, we conclude that, for the one dimensional theory under consideration, the analogue of the
exact relationship between the (co-)exterior derivative is captured by the relationship
$s_2 = +\; *\; s_1\;*$. Dimensionality of our problem also allows the validity of an inverse relationship 
(i.e. $s_1 = - * s_2 *$) between SUSY transformations $s_1$ and $s_2$.

We have discussed a
bosonic symmetry transformation $s_\omega = \{s_1, s_2 \}$ in the theory that corresponds to the Laplacian operator
$\Delta = (d + \delta)^2 = \{ d, \delta \}$. The operator form of the algebra of 
the transformations $s_1, s_2, s_\omega$ match
precisely with the algebra of the de Rham cohomological operators of differential geometry because we
have the following exact relationships, namely;
\begin{eqnarray}
&& s_1^2 = 0, \quad s_2^2 = 0, \qquad s_\omega = \{s_1, s_2 \}, \qquad [s_\omega, s_1] = 0, \qquad
[s_\omega, s_2 ] = 0, \nonumber\\
&& d^2 = 0, \qquad \delta^2 = 0, \qquad \Delta = \{ d, \delta \}, \qquad [\Delta, d] = 0, \qquad
[\Delta, \delta ] = 0.
\end{eqnarray}
Finally, we have shown, in our earlier work [14], that conserved charges of the 1D SUSY oscillator have
one-to-one correspondence with the cohomological operators of differential geometry [3-5]. We shall 
follow the logistics of our present discussion and establish the above kind of correspondence 
in the cases of potential functions which are (i) the generalizations of a harmonic oscillator
potential, and (ii) motion of a charged particle in a plane under the influence of a magnetic field
which is perpendicular to the plane.

\section{Continuous symmetries: conserved charges}
In this section, we take two different kinds of example of ${\cal N} = 2$ supersymmetric
quantum mechanical models and discuss their continuous symmetry transformations and derive the
corresponding conserved charges by exploiting the fundamental techniques of Noether's theorem.
We also establish that these conserved Noether charges are the generators of the above 
continuous and infinitesimal symmetry transformations.

\subsection{A model with the generalized SUSY potential}
We begin with the 1D general Lagrangian ($L_g$), which is a generalization of the starting Lagrangian $L_0$
[cf. (1)] with an arbitrary potential $f(x)$, as 
\begin{eqnarray}
L_g = \frac{\dot x^2 (t)}{2} - \frac{1}{2} \;\omega^2 \;\bigl (f(x) \bigr )^2 + i \;\bar\psi (t) \;\dot \psi (t) 
- \omega \; f^\prime (x)\; \bar \psi (t) \;\psi (t),
\end{eqnarray}
where $\omega$ is a parameter in the theory and $f^\prime (x) = d f/ dx$ is the first order derivative
on the potential function. It is evident that, in the limit $f (x) = x$, we retrieve our original Lagrangian
$L_0$ for the harmonic oscillator. We would like to lay stress on the fact that potential function $f (x)$
is any arbitrary (but physically well-defined) potential function and other symbols (in $L_g$) denote their
standard meanings as we have elaborated in our previous section.

The following nilpotent ($s_1^2 = 0, s_2^2 = 0$) SUSY transformations
\begin{eqnarray}
&& s_1 x = \frac{-i \;\psi}{\surd (2 \;\omega)}, \qquad s_1 \psi = 0, \qquad  s_1 \bar \psi =
\frac{1}{\surd (2 \;\omega)} [\dot x + i \;\omega \; f(x) ], \nonumber\\
&& s_2 x = \frac{i \;\bar \psi}{\surd (2 \;\omega)}, \qquad s_2 \bar \psi = 0, \qquad  s_2 \psi =
\frac{1}{\surd (2 \;\omega)} [ - \dot x + i \;\omega \;f (x) ], 
\end{eqnarray}
are the symmetry transformations for the Lagrangian $L_g$ because
\begin{eqnarray}
s_1 L_g = \;- \;\frac{d}{d t} \;\Bigl [\frac{\omega \;f(x) \;\psi}{\surd (2 \;\omega)} \Bigr ], \qquad\qquad
s_2 L_g = \;+\; \frac{d}{d t} \;\Bigl [\frac{i \;\dot x \;\bar \psi}{\surd (2 \;\omega)} \Bigr ].
\end{eqnarray}
As a consequence, the action integral ($S = \int dt\; L_g$) remains invariant under the above fermionic
transformations. The nilpotency 
properties of $s_1$ and $s_2$ is valid only on
the on-shell where the following equations of motion
\begin{eqnarray}
&& \ddot x + \omega^2 \;f\; f^\prime + \omega \; f^{\prime\prime}\; \bar \psi \;\psi = 0,  \quad
\dot \psi + i\; \omega \; f^\prime \; \psi = 0, \quad 
\dot {\bar \psi} - i\; \omega \; f^\prime \; \bar \psi = 0, \nonumber\\
&& \ddot \psi + i\; \omega\; f^{\prime\prime} \dot x\; \psi + \omega^2 \; (f^\prime)^2 \; \psi = 0, \quad
\ddot {\bar \psi} - i\; \omega\; f^{\prime\prime} \dot x\; \bar \psi + \omega^2 \; (f^\prime)^2 \; \bar\psi = 0,
\end{eqnarray}
are satisfied. The last two equations, in the above, have been derived from the basic
equations of motion $\dot \psi + i \omega  f^\prime \psi = 0, 
\dot {\bar \psi} - i \omega  f^\prime  \bar \psi = 0$. Furthermore, it is interesting to check that, under
the symmetry transformations $s_1$ and $s_2$, the above equations of motion go to one-another.

There exists a bosonic symmetry $s_\omega = \{s_1, s_2 \}$ in the theory modulo
a factor of $i$, under which, the physical variables transform as 
\begin{eqnarray}
s_\omega x = \frac{1}{\omega} \;\dot x, \quad
s_\omega \psi = \frac{1}{2\; \omega}\; \bigl (\dot \psi - i \;\omega \;f^\prime \;\psi \bigr ), \quad
s_\omega \bar \psi = \frac{1}{2 \;\omega} \;\bigl (\dot {\bar \psi} + i \;\omega \;f^\prime \;\bar \psi \bigr ).
\end{eqnarray} 
The key point to be noted here is the fact that, if we use the equations of motion, the r.h.s of
the above transformations can be written as the time derivative on the individual variables. This
verifies the existence of supersymmetry in the theory. It is a decisive feature of any arbitrary
supersymmetric theory that two consecutive supersymmetric
transformations always generate the spacetime translation. This implies that, for a 1D system,
two supersymmetric transformations should lead to the time translation
(which is satisfied in our case). Under $s_\omega$ [cf. (14)],
the Lagrangian changes as
\begin{eqnarray}
s_\omega L_g =  \frac{d}{d t} \;\Bigl [ \frac{1}{(2 \omega)}\;
\bigl (\dot x^2 - \omega^2 \; f^2 + i \;\bar \psi\; \dot \psi - \omega \;f^\prime\; \bar\psi\; \psi \bigr ) \Bigr ]. 
\end{eqnarray}
As a consequence, the action integral of our present theory remains invariant under 
the infinitesimal transformations $s_\omega = \{ s_1, s_2 \}$.

According to Noether's theorem, the above continuous symmetry transformations would lead to the derivation of 
conserved charges which would turn out to be the generators of the transformations $s_1, s_2, s_w$. To derive
these charges $(Q, \bar Q, W)$, corresponding to the above continuous symmetry transformations
($s_1, s_2, s_w$), we exploit the standard techniques and obtain the following explicit expressions
in terms of the variables of the theory:
\begin{eqnarray}
&& Q = \frac{1}{\surd (2\; \omega)} \; \Bigl [ (- i \;\dot x) \;+\; (\omega \; f(x)) \Bigr ] \; \psi \equiv
\frac{1}{\surd (2\; \omega)} \; \Bigl [ (- i \;p) \;+ \;(\omega \; f(x)) \Bigr ]\; \psi, \nonumber\\
&& \bar Q = \frac{\bar \psi}{\surd (2\; \omega)} \; \Bigl [ (+ i \;\dot x) \;+ \;(\omega \; f(x)) \Bigr ] \equiv
\frac{\bar\psi}{\surd (2\; \omega)} \; \Bigl [ (+ i \;p) \;+ \;(\omega \; f(x)) \Bigr ], \nonumber\\
&& W = \frac{1} {\omega}\; H_g \equiv \frac{1}{\omega}\; \Bigl [\frac{p^2}{2} + \frac{\omega^2\;f^2}{2}
+ \omega\; f^\prime \; \bar\psi\; \psi \Bigr ],
\end{eqnarray}
where $p = \partial L_g/\partial \dot x = \dot x$ is the canonically conjugate momentum w.r.t. the position
variable $x$ and $H_g$ is the Hamiltonian for the system under consideration. Furthermore, the equation
of motion $\dot \psi + i \omega f^\prime \psi = 0 $ has been used to express $\dot \psi$ in terms
of $\psi$ in the derivation of the Noether charge $W$. The conservation laws for these
charges can be proven by directly exploiting the equations of motion (13) and substituting them into the
expressions for 
$\dot Q, \dot {\bar Q}, \dot W$. The other way to prove the conservation laws is by computing the commutator
of the above charges with the Hamiltonian $H_g$ by exploiting the canonical brackets that emerge from the 
Lagrangian (10) of our system.

\subsection{Motion of a charged particle under influence of a magnetic field}
We consider here the well-known example of the motion of a charged particle in the $X-Y$ plane where
the magnetic field ($B_z$) is in the $Z$-direction. For the sake of simplicity, we take here 
the natural units $\hbar  = c = 1$ as well as the mass ($m$) and charge ($e$) to be unity
(i.e. $m = e = 1$). The Hamiltonian $H_{em}$ of such a charged particle, under the above magnetic field,  is [2]
\begin{eqnarray}
H_{em} = \frac{1}{2}\; (p_x + A_x)^2 + \frac{1}{2}\; (p_y + A_y)^2  - B_z \; \bar\psi\; \psi, 
\end{eqnarray}
where $A_x (x, y), A_y (x, y)$ are the components of the vector potential in the $X-Y$ plane,
$p_x = \dot x, p_y = \dot y$ are the $x$ and $y$ components of the 2D momenta  and $B_z =
\partial_x A_y - \partial_y A_x$ is the $z$-component of the magnetic field. The Lagrangian for
the above system (due to Legendre transformation) is
\begin{eqnarray}
L_{em} = \frac{1}{2}\; (\dot x^2 + \dot y^2) - (\dot x\; A_x + \dot y\; A_y) 
+ i \;\bar \psi \;\dot \psi + B_z \; \bar\psi\; \psi, 
\end{eqnarray}
where $\dot x = dx/dt, \dot y = d y/dt, \dot \psi = d \psi/dt$ are the  generalized
``velocities'' for a pair of bosonic coordinates ($x (t), y (t)$) and the fermionic variable $\psi (t)$ (in
terms of their variations w.r.t. the evolution parameter $t$). The bosonic  variables are commuting in
nature whereas the pair of fermionic variables ($\psi (t), \bar\psi (t)$) 
are anticommmuting (i.e. $\psi^2 = \bar\psi^2 = 0,
\psi \bar\psi + \bar\psi \psi = 0$).

The following  continuous and infinitesimal nilpotent ($s_1^2 = 0, s_2^2 = 0$)
fermionic  transformations ($s_1, \; s_2$):
\begin{eqnarray}
&& s_1 x = \frac{\psi}{\surd 2}, \qquad s_1 y = \frac{-i\;\psi}{\surd 2}, \qquad
s_1 \psi = 0, \qquad  s_1 \bar \psi =
\frac{i}{\surd 2} [\dot x - i \;\dot y ], \nonumber\\
&& s_1 A_x = \frac{1}{\surd 2} \bigl (\partial_x A_x - i\; \partial_y A_x \bigr ) \; \psi, \qquad
s_1 A_y = \frac{1}{\surd 2} \bigl (\partial_x A_y - i \;\partial_y A_y \bigr )\; \psi, \nonumber\\ 
&& s_2 x = \frac{\bar \psi}{\surd 2}, \qquad s_2 y = \frac{i\;\bar \psi}{\surd 2}, \qquad
s_2 \bar \psi = 0, \qquad  s_2  \psi =
\frac{i}{\surd 2} [\dot x + i \;\dot y ], \nonumber\\
&& s_2 A_x = \frac{\bar \psi}{\surd 2} \bigl (\partial_x A_x + i \;\partial_y A_x \bigr ), \qquad
s_2 A_y = \frac{\bar \psi}{\surd 2} \bigl (\partial_x A_y + i \;\partial_y A_y \bigr ), 
\end{eqnarray}
are the symmetry transformation for the action integral ($S = \int dt L_{em}$) because 
the Lagrangian (18) transforms to the total derivatives as 
\begin{eqnarray}
&& s_1 L_{em} = \;- \;\frac{d}{d t} \;\Bigl [\frac{(A_x - i\; A_y) \;\psi}{\surd (2)} \Bigr ], \nonumber\\
&& s_2 L_{em} = \;+\; \frac{d}{d t} \;\Bigl [\frac{\bar \psi}{\surd (2)} \;
\bigl \{\dot x + i \dot y - (A_x + i A_y) \bigr \} \Bigr ].
\end{eqnarray}  
Thus, according to Noether's theorem, we shall have conserved charges which would turn out to be the generators
for the above continuous symmetries. These SUSY fermionic ($ Q^2 = \bar Q^2 = 0$) charges, corresponding to the nilpotent SUSY transformations $s_1$ and $s_2$, are as follows
\begin{eqnarray}
&& Q = \frac{1}{\surd (2\;)} \; \Bigl [ (p_x + A_x) - i \;(p_y + A_y) \Bigr ] \; \psi, \nonumber\\
&& \bar Q = \frac{\bar \psi}{\surd (2)} \; \Bigl [ (p_x + A_x) + i \;(p_y + A_y) \Bigr ], 
\end{eqnarray}
which are derived using the standard techniques of Noether's theorem.

The anticommutator $\{ s_1, s_2 \} = s_\omega$ leads to the derivation of a bosonic symmetry in the theory. 
The continuous and infinitesimal version of these transformations (modulo a factor of $i$) 
on the generic variable $\Phi$ is 
\begin{eqnarray}
s_\omega \; \Phi =  \; \dot \Phi, \qquad  \Phi = x (t),\; y (t), \;\psi (t),\; \bar \psi (t),\;
 A_x (x, y),\; A_y (x, y).
\end{eqnarray}
In the derivation of the above bosonic symmetry transformations, for obvious reasons, we have used
the following straightforward inputs, namely;
\begin{eqnarray}
&& \partial_x \psi (t) = 0, \qquad \partial_y \psi (t) = 0, \qquad 
\partial_x \bar \psi (t) = 0, \qquad \partial_y \bar\psi (t) = 0, \nonumber\\
&& \frac{d}{dt} A_x (x, y) = \dot x \; \partial_x A_x + \dot y\; \partial_y A_x, \quad
\frac{d}{dt} A_y (x, y) = \dot x \; \partial_x A_y + \dot y\; \partial_y A_y.
\end{eqnarray}
The Lagrangian of our present theory transforms to a total derivative rendering the action integral invariant.
Mathematically, this statement is
\begin{eqnarray}
s_\omega \; L_{em}  = \frac{d}{dt} \Bigl [ L_{em} \Bigr ] \;\;\;\;\;\Rightarrow\;\;\;\; 
s_\omega S = \int \; dt \;\bigl (s_\omega L_{em} \bigr ) = 0.
\end{eqnarray}
Applying the standard techniques of Noether's theorem, we obtain the expression of the conserved charge
as follows
\begin{eqnarray}
W =  \; H_{em}  \equiv \; \Bigl [\frac{(p_x + A_x)^2}{2} +  \frac{(p_y + A_y)^2}{2}
- B_z\; \bar\psi\; \psi \Bigr ].
\end{eqnarray}
Thus, we note that the conserved charge, corresponding to the bosonic symmetry transformations, is nothing  but
the Hamiltonian of the theory itself. The conservation laws ($\dot Q = \dot {\bar Q}
= \dot W = 0$) can be proven by using the equations of motion that emerge from $L_{em}$. There is another
way to prove the conservation laws, too. One can compute the commutator of the Hamiltonian with the charges
$Q, \bar Q, W$ by exploiting the canonical (anti)commutators that are deduced 
from the Lagrangian $L_{em}$ [cf. (38) and Appendix, below].

We wrap up this section with a couple of general remarks. First, the conserved charges in (16), (21) and
(25)  are the generators of the infinitesimal transformations $s_1, s_2, s_\omega$ 
because we have the following relationships
\begin{eqnarray}
s_r \Phi = \mp\; i\; [\Phi, \;Q_r ]_{(\pm)}, \qquad \; s_r = s_1, s_2, s_\omega, \qquad \; Q_r = Q, \bar Q, W,
\end{eqnarray}
where the $(+)-$ signs, as the subscripts on the square bracket, stand for the bracket to be (anti)commutator for the generic variable $\Phi = x, \psi, \bar\psi$ and
$\Phi = x, y, \psi, \bar\psi, A_x, A_y$ being (fermionic) bosonic in nature
for both the SUSY examples of our present endeavor. The $(+)-$ signs, in front of the square bracket, are chosen judiciously. A detailed discussion about the choice of the latter can be found in our earlier work
(see, e.g. [16]). Second, it is clear that the fermionic variables $\psi$ and $\bar \psi$ remain invariant
under the nilpotent transformations $s_1$ and $s_2$, respectively. As pointed out in equation (4), there is a duality
symmetry (i.e. discrete symmetry) in the theory. One can argue that the nilpotent symmetry transformations $s_1$ and $s_2$ are {\it dual} to each-other because they leave $\psi$ and $\bar\psi$ invariant which are connected to each-other
by the duality transformations in (4) and its generalized form (see below).

\section{Discrete symmetries: duality transformations}
In this section, we discuss the presence of a set of discrete symmetry transformations for both the
${\cal N} = 2$ supersymmetric quantum mechanical models under consideration and establish their relevance 
to the Hodge duality $(*)$ operation of differential geometry.

\subsection{A model with the generalized SUSY potential}
It is very interesting to note that there is a set of discrete symmetries in the theory because the Lagrangian $L_g$ [cf. (10)] remains invariant under these specific transformations. 
Let us focus on the discrete transformations
\begin{eqnarray}
&& x \to  - \;x, \qquad \omega \to - \;\omega, \qquad \psi \to \pm \;i \;\bar \psi, \qquad
\bar\psi \to \mp \;i \;\psi,\nonumber\\
&& t \to t, \qquad \;\;f (x) \to - f (x), \qquad \;\;\;f^\prime (x) \to f^\prime (x),
\end{eqnarray}
where we have denoted the 
actual transformations: $x \to x^\prime = - x, f (x) \to f (- x) = - f(x)$, etc., in an abbreviated
form. Furthermore, as is evident,  there are two symmetry transformations that are hidden in the above transformations. The discrete 
 transformations (27) are actually the generalization of transformations in (4) because, in the limit $f (x) = x$, we
retrieve (4) from (27). It is clear that, physically, we are talking about the parity 
transformation operator ($\hat P$), under which,
the potential function $f(x)$ has to be {\it odd} because we are 
theoretically compelled to choose $f^\prime (-x) = f^\prime (x)$  
to incorporate the useful 
discrete symmetry transformations in our present theory.

In the case where the first-order derivative on the potential function is 
chosen to be {\it odd} under the parity operator
(i.e. $ \hat P f^\prime (x) \equiv f^\prime (- x) = - f^\prime (x)$), we 
are theoretically forced to rely on the following transformations
\begin{eqnarray}
&& x \to  - \;x, \qquad \omega \to - \;\omega, \qquad \psi \to \pm \;i \;\bar \psi, \qquad
\bar\psi \to \pm \;i \;\psi,\nonumber\\
&& t \to - t, \qquad \;\;f (x) \to  f (x), \qquad \;\;\;f^\prime (x) \to - f^\prime (x),
\end{eqnarray}
under which, the Lagrangian $L_g$ remains invariant. A close look at the above transformations implies, physically,
that there is time-reversal ($\hat T$) as well as parity 
($\hat P$) invariance in the theory where the potential function  is an
even function under parity (i.e., $\hat P f(x) = + f(x)$). As a consequence, the first-order derivative on the potential
function, automatically, becomes an odd function of parity (i.e. $ \hat P f^\prime (x) = - f^\prime (x)$. Furthermore, we note that the fermionic variables transform, under the time-reversal operator, as
\begin{eqnarray}
\hat T: t \to - t, \qquad \hat T\; \psi (t) = \pm i\; \bar \psi (t), \qquad 
\hat T \;\bar \psi (t) = \pm i\;  \psi (t).
\end{eqnarray}
We observe, in passing, that the generalized velocity $\dot x$ is an invariant quantity under
the combined operations of parity and time-reversal as is evident from $\hat T \hat P \;(\dot x) = + (\dot x)$.
Physically, this observation shows that the kinetic term for the bosonic variable of our present supersymmetric
system is $PT$-invariant. However, the kinetic term 
(i.e. $i \;\bar\psi \;\dot \psi$), for the fermionic part of our present model, for obvious reasons,
is time-reversal invariant (i.e. $\hat T\; (i\;\bar\psi \;\dot \psi) = i\; \bar\psi\; \dot \psi$).

The Lagrangian $L_g$, with the general potential function $f (x)$, has also {\it only} time-reversal symmetry.
It is elementary to check that the following transformations (corresponding to the time-reversal operator), namely;
\begin{eqnarray}
&& x \to  + \;x, \qquad \omega \to + \;\omega, \qquad \psi \to \pm \;i \;\bar \psi, \qquad
\bar\psi \to \pm \;i \;\psi,\nonumber\\
&& t \to - t, \qquad \;\;f (x) \to  f (x), \qquad \;\;\;f^\prime (x) \to  f^\prime (x),
\end{eqnarray}
leave the Lagrangian $L_g$ invariant. It is to be pointed out that, in the above, there is no
reflection symmetry in the theory because $x \to x$. As a consequence, the potential function
as well as its first-order derivative remain unaffected due to the presence of time-reversal symmetry
{\it alone} [cf. (18)].

We dwell a bit now on the importance of the discrete symmetry transformations we have discussed so far. The
discrete symmetry transformations (27) [that are generalization of (4)] correspond to the Hodge duality
$*$ operation of differential geometry. This can be proven by checking that relations (5) are satisfied by the
interplay of continuous and discrete symmetry transformations (11) and (27) when they blend together in a
meaningful manner. Furthermore, we find that relation (6) is also valid in the case of general potential 
function $f(x)$ for our {\it first} ${\cal N} = 2$ SUSY example. As a consequence, we observe that relations
(7) and (9) are also true. Thus, it is crystal clear that the
relationship (7) is the analogue of the relationship that exists between the (co-)exterior derivatives 
[$(\delta)d$] of differential geometry.

Now we comment on the existence of discrete symmetry transformations (28) and (30), under which,
the Lagrangian $L_g$ remains invariant, too. It turns out that neither set of these symmetries leads to
the exact derivation of relationship like (7) and its counterpart $s_1 = - * s_2 *$. As a consequence, these symmetries are {\it not} interesting
from the point of view of the duality-invariant physical theories [15]. To elaborate on this statement, first
of all, we note
that the transformations (30) {\it outrightly} do not yield $s_2 = \pm \;*\;s_1\;*$. Furthermore, two successive
operations of (28) produces: $*\;(*\;x) = + x, \;*\;(*\;\Phi) = - \Phi$ where $\Phi = \psi, \bar\psi$. As a
consequence, we have the relationships $s_2 x = + * s_1  * x$ and $s_2 \Phi = - * s_1 * \Phi$ for 
$\Phi = \psi, \bar\psi$. These are very nicely satisfied by (28). However, the reverse relationships
$s_1 x = - * s_2 * x$ and $s_1 \Phi = + * s_2 * \Phi$ are {\it not} satisfied by the discrete symmetry
transformations (28). Thus, we ignore (28) as the physical realization of the Hodge duality 
$(*)$ operation.

We emphasize that, ultimately, the following 
{\it unique} transformations
\begin{eqnarray}
&& x \to  - \;x, \qquad \omega \to - \;\omega, \qquad \psi \to + \;i \;\bar \psi, \qquad
\bar\psi \to - \;i \;\psi,\nonumber\\
&& t \to t, \qquad \;\;f (x) \to - f (x), \qquad \;\;\;f^\prime (x) \to f^\prime (x),
\end{eqnarray}
correspond to the Hodge duality $*$ operation of differential geometry. At this stage, there 
are a couple of remarks. First, it is physically
very important that the potential function turns out to be odd under parity (see, e.g. [17,2] for details).
Second, it can be checked that the reverse relationship ($s_1 = -\;*\;s_2\;*$) of (7) also exists in 
the theory because of its dimensionality.
Henceforth, we shall concentrate on
the above unique transformations as the analogue of the Hodge duality $*$ operation (as far as the physical discussions
of our present theory, with a general super potential function $f(x)$, is concerned).

We close this section with the remark that the super charges $Q$ and $\bar Q$ transform under the
the duality transformations (31) as follows
\begin{eqnarray}
* \;(Q) = \bar Q, \qquad * \; (\bar Q) = - Q, \qquad *\; (*\; Q) = - Q, \qquad *\; (*\; \bar Q) = - \bar Q.
\end{eqnarray}
We point out that $Q$ and $\bar Q$ transform in exactly same manner as the 
electric-magnetic duality transformations for the source-free Maxwell's equations where 
${\bf E} \to {\bf B}, {\bf B} \to - {\bf E}$. Thus, there is a perfect duality symmetry in our theory.
Furthermore, we note that, in contrast to the transformations (6), we find that the double $*$ operations
on the fermionic charges results in a negative sign. Another interesting observation is:
$* \; W = - W,\; *\; (*\; W) = + W$. As a consequence, we find that the $sl(1/1)$ algebraic
structure: $Q^2 = 0, \bar Q^2 = 0, \{Q, \bar Q\} = (H_g/\omega)$
remains invariant under the $*$ duality transformations (applied any arbitrary number of times on this algebra). Finally, we note
that the Hamiltonian of the theory remains duality invariant (because $*\; H_g = H_g$) 
as is the case with the Lagrangian ($ *\; L_g = L_g$) of our present theory.\\

\subsection{Motion of a charged particle under influence of a magnetic field}
Unlike the previous subsection 4.1, we shall focus here {\it only} on those discrete symmetry transformations
which are useful to us as far as the derivation  of
connection between SUSY transformations $s_1$ and $s_2$ is concerned.
In fact, it can be checked that the Lagrangian $L_{em}$ remains invariant under the following useful discrete
symmetry transformations
\begin{eqnarray}
&& x \to  \mp \;x,  \qquad \psi \to \mp \;\bar \psi, \qquad A_x \to \pm\; A_x, \qquad t \to -\; t, \nonumber\\
&& y \to \pm\; y, \qquad \bar\psi \to \pm  \;\psi, \qquad A_y \to \mp\; A_y, \qquad B_z \to B_z.
\end{eqnarray}
A few remarks are in order at this stage. First,
it should be noted that, in reality, there are two discrete transformations in (33) that leave the Lagrangian
$L_{em}$ invariant. Second, there is always a time reversal ($ t \to - t$) symmetry in the theory irrespective
of how the coordinates $(x, y)$, in the plane, transform. Third, as a consequence of transformations
($x \to \mp x, y \to \pm y$), the space derivatives transform as $\partial_x \to \mp \partial_x,
\partial_y \to \pm \partial_y$. Fourth, the kinetic terms $(\dot x^2/2)$ and $(\dot y^2/2)$
remain invariant under (33).
Finally, we would like to remark that the vector potentials change explicitly,
under the parity-type transformations for the space variables ($x \to \mp x, y \to \pm y$), as
\begin{eqnarray}
&& A_x (x, y) \to  A_x (\mp \;x, \pm \;y) \;= \; \pm\; A_x (x, y),\nonumber\\
&& A_y (x, y) \to  A_y (\mp \;x, \pm \;y) \;= \; \mp\; A_y (x, y),
\end{eqnarray}
which leave the magnetic field $B_z = \partial_x A_y - \partial_y A_x$ invariant (because we have to take into account
the corresponding transformations for the derivatives $\partial_x \to \mp \partial_x, \partial_y \to \pm \partial_y$
under the above transformations $x \to \mp x, y \to \pm y$).

We would like to re-state the following.
As discussed in subsection 4.1, we can also pay attention to various possibilities of the existence of
discrete symmetries in our present model of ${\cal N} = 2$ supersymmetric quantum mechanical system. However,
we have concentrated {\it only} on the symmetry transformations 
(33) that are very useful to us as far as the derivation
of the analogue of the relationship $\delta = \pm\; *\; d\; *$ (in the language of symmetry transformations)
is concerned. In fact, as it turns out, we shall see that it is the interplay of the continuous nilpotent
($s_1^2 = 0, s_2^2 = 0$) transformations $s_1, s_2$ and
discrete symmetry transformations (33) that provide the analogue of the relationship between the (co-)exterior
derivatives ($\delta, d$) of differential geometry.

To appreciate the importance of the discrete symmetry transformations (33), first of all, we observe that
two successive operations of these discrete symmetry transformations on any generic variable $\Phi$ yields
either plus or minus sign. This can be mathematically stated as $*\; [* \Phi] = \pm \; \Phi$ where the
$*$ operation 
is nothing but the discrete symmetry transformations (33) and the generic variable $\Phi = x, y, \psi, \bar\psi,
A_x, A_y$. To be more specific, it can be seen that the following is true (for $\Phi_1$ and $\Phi_2$
components of $\Phi$), namely;
\begin{eqnarray}
&& *\; [\; *\; ] \; \Phi_1 = \; +\; \Phi_1,  \qquad \qquad \Phi_1 = x, y, A_x, A_y, \nonumber\\
&& *\; [\; *\; ] \; \Phi_2 = \; -\; \Phi_2,  \qquad \qquad \Phi_2 = \; \psi , \; \bar \psi.
\end{eqnarray} 
The connection between the (co-)exterior derivatives $(\delta)d$ (i.e. $\delta = \pm\; *\; d \; *$) can
be realized between the nilpotent fermionic symmetry transformations $s_1$ and $s_2$ (i.e. $s_2 = \pm *\; s_1 \:*$).
To pin-point this relationship in a more specific fashion
(see, e.g. [15]), we have the following explicit relationships, namely;
\begin{eqnarray}
&& s_2 \Phi_1 = +\; * s_1\; *\; \Phi_1 \;\;\Rightarrow\;\; s_2 = + \;*\; s_1\; *,  \nonumber\\
&& s_2 \Phi_2 = -\; * s_1\; *\; \Phi_2 \;\;\Rightarrow\;\; s_2 = - \;*\; s_1\; *,
\end{eqnarray} 
where, as is evident, $\Phi_1 = x, y, A_x, A_y$ and $\Phi_2 = \psi, \bar\psi$. We note that
the reverse relationships $s_1 \Phi_1 = - * s_2 * \Phi_1$ and $s_1 \Phi_2 = + * s_2 * \Phi_2$ are also
true.

As a final remark, we mention here that the Lagrangian and Hamiltonian of our present model are duality invariant
(i.e. $*\; L_{em} = L_{em}, \; *\; H_{em} = H_{em}$) and
the fermionic conserved charges transform under (33) as
\begin{eqnarray}
 *\; Q = \mp\; \bar Q, \quad *\; \bar Q = \pm\; Q, \quad
*\; [\; *\;  Q] \; = \; -\; Q,  \qquad 
*\; [\; *\; \bar Q] = -\; \bar Q.
\end{eqnarray} 
Thus, once again, we observe the duality transformations (i.e. ${\bf E} \to \pm {\bf B}, {\bf B} \to \mp {\bf E}$)  
of source-free Maxwell equations being replicated here in the duality transformations of $Q$ and $\bar Q$. Furthermore,
we note that the $sl(1/1)$ closed superalgebra here does {\it not} remain invariant under the duality $*$ operation
because the Hamiltonian turns out to be duality invariant (i.e. $*\; H_{em} = H_{em}$). This discrepancy
(from the model of our previous subsection 4.1) appears because of the fact that parameter $\omega$ is 
{\it absent} in our present model. This is the reason that, in our case, we have  $*\; W = + W$ 
(as $ W = H_{em}$). On the contrary, in our previous subsection, we had $*\; W = -\; W$ 
because $*\; (H/\omega) = - (H_g/\omega)$ (due to $*\; H_g = H_g$ and $ *\;\omega = - \omega$).

\section{Algebraic structures: cohomological aspects}
In the present section, we shall discuss the algebraic structures
of the conserved charges ($Q, \bar Q, W$) for both the ${\cal N} = 2$ SUSY quantum
mechanical models {\it together} that have been considered in our present endeavor.

We have already noted that the fermionic SUSY transformations $s_1$ and $s_2$ are nilpotent
of order two (i.e. $s_1^2 = s_2^2 =0 $) on the on-shell where the Euler-Lagrange equations of motion
(13) are satisfied for our {\it first} model of SUSY example. In the case of the motion of a charged particle,
we observe that the nilpotency of $s_1$ and $s_2$ ensue from the fermionic (i.e. $\psi^2 = \bar\psi^2 = 0$)
nature of variables $\psi$ and $\bar\psi$.
Furthermore, it can be explicitly checked that the bosonic symmetry transformation
$s_\omega$, that is equal to the anticommutator (i.e. $s_\omega = \{s_1, s_2\}$) of the fermionic transformations,  commutes with both the  SUSY transformations $s_1$ and $s_2$. As a consequence, the operator form of the
transformations $s_\omega$ is the Casimir operator for the whole algebra. Thus, we conclude that the
operators $s_1, s_2, s_\omega$ satisfy exactly the same algebra as is the case of
SUSY harmonic oscillator [cf. (9)].

It turns out that the conserved charges of (16), (21) and (25)
obey exactly the same algebra as the operator form of the transformations $s_1, s_2, s_\omega$. Mathematically,
this super algebra $sl(1/1)$ can be succinctly written  as
\begin{eqnarray}
Q^2 = 0 \quad \bar Q^2 = 0, \quad W =  \{Q, \;\bar Q \}, \quad [W, \;Q] = 0, \quad [W, \;\bar Q] = 0.
\end{eqnarray}
In view of the fact that $W = (H_g/\omega)$ and $W = H_{em}$ for both the SUSY models, respectively,
it is obvious that the last two entries in  the above
equation are nothing but the conservation laws (i.e. $\dot Q = \dot {\bar Q} = 0$) for  $Q$ and $\bar Q$.
Furthermore, it is crystal clear that the conserved bosonic charge $W$ is the Casimir
operator for the whole algebra. A close look at (38) shows that its algebraic structure is exactly same as
the algebraic structure of the de Rham cohomological operators of differential geometry [(cf. (9)].

Due to the above observations, it is very tempting to identify the set of conserved charges 
$(Q, \bar Q, W)$ with the set of cohomological operators $(d, \delta, \Delta)$ of differential geometry.
However, the identification is {\it not} yet complete 
because the cohomological operators satisfy specific properties 
when they operate on the differential form of a definite degree. For instance, it is a well-known fact that the (co-)exterior derivatives (lower)raise the degree of a form by {\it one} when they operate on it. On the contrary, the
Laplacian operator does not change the degree of the form on which it acts. We have to capture
these properties in the language of conserved charges (i.e. $Q, \bar Q, W$)
 for the completion and correctness of an exact identification.

To achieve the above goal, we have taken the help of bosonic as well as fermionic number operators 
(and their eigen-values) in the context of SUSY quantum mechanical harmonic oscillator 
where $ f(x) = x$ [14]. For an arbitrary potential function $f (x)$, the above arguments fail 
because the bosonic creation and annihilation operators become non-trivial and their commutation
relation produce a first-order derivative $f^\prime (x)$ on the potential function $f(x)$. In exactly
similar fashion, the 
{\it second} example of our ${\cal N} = 2$ SUSY model (connected with the motion of a charged particle)
also does not obey the above logic.
However, it is illuminating to note that the following algebra, amongst 
the set of operators ($Q, \bar Q, W)$, is true, namely;
\begin{eqnarray}
&& [Q \; \bar Q, \;Q ] = + W \; Q, \qquad \qquad [Q \; \bar Q, \;\bar Q ] = - W \; \bar Q, \nonumber\\
&& [\bar Q \; Q,\; Q ] = - W \; Q, \qquad \qquad  [\bar Q \; Q, \;\bar Q ] = + W \; \bar Q,
\end{eqnarray}
where, as is evident from (38), the charge $W = \{Q, \bar Q\}$ is 
the Casimir operator  (i.e. $ [W,\; Q] = [W, \;\bar Q] = 0$).
We assume that the inverse of the Casimir operator ($W^{-1}$) is well-defined and the latter {\it logically} commutes
with both the nilpotent super charges (i.e. $ [W^{-1}, Q] = [W^{-1}, \bar Q] = 0$).

As a consequence of the above arguments, the algebra (39)
can be re-expressed, in a theoretically useful and handy manner, as follows 
\begin{eqnarray}
&& \Bigl [\frac{Q \; \bar Q}{W}, \;Q \Bigr ] = + \; Q, \qquad \qquad 
\Bigl [\frac{Q \; \bar Q}{W}, \;\bar Q \Bigr ] = -  \; \bar Q, \nonumber\\
&& \Bigl [\frac{\bar Q \; Q}{W},\; Q \Bigr ] = - \; Q, \qquad \qquad 
\Bigl [\frac{\bar Q \; Q}{W}, \;\bar Q \Bigr ] = +  \; \bar Q,
\end{eqnarray}
In this situation, one can
define a state $|\chi>_p$, in the quantum Hilbert space of states (QHSS), which satisfies 
$(Q \bar Q/W)\; |\chi>_p = p \; |\chi>_p$ where $p$ is the eigen-value of operator 
$(Q \bar Q/W)$. Using the top two relations of (40), it can be checked that
the states $Q\; |\chi>_p, \bar Q \; |\chi>_p, W\; |\chi>_p$ satisfy 
\begin{eqnarray}
\Bigl (\frac{Q \;\bar Q}{W} \Bigr )\;  Q\; |\chi>_p &=& (p + 1) \; Q\; |\chi>_p, \nonumber\\
 \Bigl (\frac{Q \;\bar Q}{W} \Bigr )\;  \bar Q\; |\chi>_p &=& (p - 1) \; \bar Q\; |\chi>_p, \nonumber\\
 \Bigl (\frac{Q \;\bar Q}{W} \Bigr )\;  W\; |\chi>_p &=& (p) \; W\; |\chi>_p.
\end{eqnarray}
As a consequence, we note that the states $Q\; |\chi>_p, \bar Q \; |\chi>_p, W\; |\chi>_p$ have the eigen-values 
$(p + 1), (p - 1), (p)$, respectively. This establishes the fact that if the degree of a form 
is identified with the eigen-value of a specific state in the QHSS for the operator $(Q \bar Q/W)$, the 
result of the operation of conserved charges $(Q, \bar Q, W)$ on this particular state is exactly same
as the consequences that follow after
the operation of the cohomological operators $(d, \delta, \Delta)$ on the specific degree of a form
(which is equal to the above eigen-value).
Thus, ultimately, we have the following one-to-one mapping        
\begin{eqnarray}
(Q, \; \bar Q, \; W) \qquad \Leftrightarrow \qquad (d, \; \delta,\; \Delta),
\end{eqnarray}
between the conserved charges corresponding to the physical symmetries of the theory and the cohomological
operators of differential geometry.

Now we exploit the lower two relations of (40) and define an arbitrary state $|\xi>_q$ to possess
the eigen-value $q$ w.r.t. the operator $(\bar Q\; Q)/W$ [i.e. $(\bar Q\; Q)/W\; |\xi>_q = q\; |\xi>_q$].
In view of this definition, the following theoretically interesting relationships automatically ensue
\begin{eqnarray}
\Bigl (\frac{\bar Q \;Q}{W} \Bigr )\;  Q\; |\xi>_q &=& (q - 1) \; Q\; |\xi>_q, \nonumber\\
\Bigl (\frac{\bar Q \;Q}{W} \Bigr )\;  \bar Q\; |\xi>_q &=& (q + 1) \; \bar Q\; |\xi>_q, \nonumber\\
\Bigl (\frac{\bar Q\; Q}{W} \Bigr )\;  W\; |\xi>_q &=& (q) \; W\; |\xi>_q.
\end{eqnarray}
The above relationships establish that the states $Q\; |\xi>_q, \bar Q \; |\xi>_q, W\; |\xi>_q$
have the eigen-values $(q - 1), (q + 1), (q)$, respectively. Thus, we conclude that if the degree
of a form is identified with the eigen-value $q$ of a state in the QHSS corresponding to the operator
$(\bar Q\; Q)/W$, there is one-to-one relationship between the conserved charges ($\bar Q, Q, W$)
corresponding to the continuous symmetries of the theory {\it and} the cohomological operators:
\begin{eqnarray}
(\bar Q, \; Q, \; W) \qquad \Leftrightarrow \qquad (d, \; \delta,\; \Delta),
\end{eqnarray} 
as far as the analogy between the eigen-values and the degree $q$ of a given form is concerned. Thus,
we have proven that our present couple of ${\cal N} = 2$ SUSY models 
are very interesting physical models for the Hodge theory where all the de Rham cohomological
operators, Hodge duality operation, degree of a form, etc., find their physical realizations
in the language of discrete and continuous symmetry transformations (and corresponding generators).

\section {Conclusions} 
In our present investigation, we have shown that a triplet of well-known SUSY quantum mechanical
systems are tractable models for the Hodge theory. We have touched very briefly upon the proof that
the 1D SUSY harmonic oscillator is a model for the Hodge theory. An extensive 
discussion on this observation can be found in [14]. We have provided definite
proofs, however, for the other two ${\cal N} = 2$ SUSY systems of our present investigation
and demonstrated that these systems
are also models for the Hodge theory.
We conjecture, in our present endeavor, that any arbitrary ${\cal N} = 2$ SUSY quantum
mechanical model could be shown to respect continuous and discrete symmetries  that are physical realizations of
the de Rham cohomological operators and Hodge duality operation
of differential geometry, respectively. As a consequence, the above
set of ${\cal N} = 2$ SUSY models are very special.

All the above SUSY models are endowed with two SUSY transformations ($s_1, s_2)$ and a 
bosonic symmetry transformation $s_\omega$. In our present investigation, we have defined the
latter symmetry as an anticommutator of the above two SUSY transformations modulo
a factor of $i$ because $s_\omega$ corresponds to the Laplacian operator which is,
as is well-known, a hermitian operator with a positive real eigen-value [3-5]. In fact,
it is because of the above choice that the conserved charge $W$ (which is the generator
of the bosonic symmetry transformation $s_\omega$) turns out to be hermitian
(i.e. $W = (H_g/\omega)$ and $W = H_{em}$) for both the ${\cal N} = 2$ SUSY models under consideration. Furthermore,
the above observation (at the symmetry level) is also reflected in the $sl(1/1)$
algebra satisfied by the conserved charges $(Q, \bar Q, W)$ which are the generators
of the continuous symmetry transformations $(s_1, s_2, s_\omega)$.

We observe that, for the 1D system of ${\cal N} = 2$ SUSY model, we obtain only one discrete
symmetry transformation that is consistent with the strictures laid down by the duality-invariant
physical theories [15]. As a consequence, we have only one relationship between the SUSY transformations
$s_1$ and $s_2$ (i.e. $s_2 = + * s_1 *$) as an analogue of the well-known connection between
the (co-)exterior derivatives: $\delta = \pm *  d *$. On the contrary, for the 2D case of the
motion of a charged particle (corresponding to our {\it second} example of ${\cal N} = 2$ SUSY model),
we have a set of two discrete symmetries [cf. (33)]. As a result, we have two relationships  $s_2 = \pm * s_1 *$
that are precise analogues of the relationships between the (co-)exterior derivatives: $\delta = \pm * d *$.
In addition to the above observations, we note that there is always a time-reversal 
($t \to - t$) discrete symmetry in the case of 2D ${\cal N} = 2$ SUSY theory [cf. (33)] which is not present
in the case of 1D ${\cal N} = 2$ SUSY model of our present investigation [which is clear from equation  (31)].

It is an open question as to why there is only one physically consistent discrete symmetry
for the 1D ${\cal N} = 2$ SUSY model whereas there are two physically consistent discrete symmetry 
transformations for the 2D model of ${\cal N} = 2$ SUSY system. In our earlier works
on Abelian $p$-form ($ p = 1, 2, 3...$) gauge theories (within the framework of BRST formalism)
[6-10], we have established that
such theories are examples of Hodge theories when the spacetime dimension $D$ is equal to $2 p$
(i.e. $D = 2 p$). In these theories, we have shown the existence of two physically important
discrete symmetry transformations. We do not know, at the moment,  whether there is any type of connection
between the SUSY theories and gauge theories (as far as theoretical aspects of
models for the Hodge theory are concerned). These are some of the issues that we plan to address
in our future investigations.

\section*{Acknowledgments}
One of us (RPM) is grateful to the Director, IISER, Pune, for the warm
hospitality extended to him during his visit to the THEP group (of IISER) where a part of this
work was completed. Fruitful comment by our esteemed Referee is gratefully acknowledged, too.

\appendix
\section{Simpler ways of deriving $sl(1/1)$}

Here we discuss the simpler ways of deriving the closed super algebra $sl(1/1)$
amongst the conserved charges $(Q, \bar Q, W)$ by exploiting the canonical
definition of the generator of a continuous symmetry transformation [cf. (26)]. For
the first example of ${\cal N} = 2$ SUSY model with the generalized potential function $f (x)$,
we can exploit the nilpotent ($s_1^2 = s_2^2 = 0$)
symmetry transformations (11) to compute the l.h.s. of the following equation
\begin{eqnarray}
s_1 \; \bar Q = i\; \{\bar Q, Q \} \equiv i \; W, \qquad \qquad
s_2 \;  Q = i\; \{Q,  \bar Q \} \equiv i \; W.
\end{eqnarray}
With the inputs from (16) for the expressions of charges, we can
demonstrate that the l.h.s. matches with the r.h.s. with $i \; W = i \;(H_g/\omega)$. Thus, we derive
the relationship $\{ Q, \bar Q\} = (H_g/\omega) \equiv W$.
The whole beauty of this simple derivation is the mere use of (11) and (16) in the derivation
of {\it one} of the most important ingredients of the $sl(1/1)$ superalgebra.

To prove the nilpotency (i.e. $Q^2 = \bar Q^2 = 0$) of the super charges $ Q$ and $\bar Q$, 
we exploit the following appropriate relationships
\begin{eqnarray}
s_1 \; Q = i\; \{Q, Q \} \equiv 0, \qquad \qquad
s_2 \;  \bar Q = i\; \{\bar Q,  \bar Q \} \equiv 0,
\end{eqnarray}
where, once again, the SUSY transformations $s_1$ and $s_2$ from (11) and
expressions for the charges $Q$ and $\bar Q$ from (16) have been used in the evaluation
of the l.h.s. of the above relationships. The above equations (45) and (46) show the
validity and deduction of $sl(1/1)$ closed super algebra (38) amongst the conserved
nilpotent charges $Q, \bar Q$ and the bosonic charge $W = (H_g/\omega)$.

We wrap up this Appendix with the remarks that the analogues of computations (45) and (46)
can be performed for the {\it second} ${\cal N} = 2$ SUSY example of the motion of a charged particle
under influence of a magnetic field where the nilpotent transformations (19) and expressions
for the charges in (21) and (25) can be exploited for the evaluation of variations
$s_1 \bar Q$ and $s_2 Q$ which lead to the derivation of $\{ Q, \bar Q \} = W$ where
$W$ turns out to be equal to the Hamiltonian $H_{em}$ (i.e. $W = H_{em}$). 
Similarly, the nilpotency of the
charges $Q$ and $\bar Q$ [cf. (21)] can be proven by exploiting the nilpotent transformations
(19). In other words, we evaluate $s_1 Q$ and $s_2 \bar Q$ which turn out to
yield $Q^2 = \bar Q^2 = 0$. We wish to lay emphasis on the fact that it is the definition
of the generator of a continuous symmetry transformation [cf. (26)] that plays a key
role in the derivation of the superalgebra $sl(1/1)$.





\bibliographystyle{model1a-num-names}
\bibliography{<your-bib-database>}







\end{document}